\documentclass[11pt,a4paper]{article}

\usepackage{amssymb, amsmath} 
\usepackage{fullpage}
\usepackage{apacite}
\usepackage{natbib} 

\usepackage{setspace}
\begin{document}

\date{July 12,  2013}
\title{Commentary on Quantum-Inspired Information Retrieval}

\author{
        Elham Ashoori\\
                Controlled Quantum Dynamics Theory Group\\
								Imperial College London, London SW7 2AZ, United Kingdom\\
								e.ashoori@imperial.ac.uk\\
            \and
        Terry Rudolph\\
        Controlled Quantum Dynamics Theory Group\\
								Imperial College London, London SW7 2AZ, United Kingdom\\
								tez@imperial.ac.uk
}

\maketitle

\clearpage

\begin{center}
\setlength{\topskip}{1in}
{\LARGE{Commentary on Quantum-Inspired Information Retrieval}}
\end{center}

\vspace{10 mm}

\begin{abstract}
There have been suggestions within the Information Retrieval~(IR) community that quantum mechanics (QM) can be used to help formalise the foundations of IR.
The invoked connection to QM is mathematical rather than physical. The proposed ideas are concerned with information which is encoded, processed and accessed in classical computers. However, some of the suggestions have been thoroughly muddled with questions about applying techniques of quantum information theory in IR, and it is often unclear whether or not the suggestion is to perform actual quantum information processing on the information.
This paper is an attempt to provide some conceptual clarity on the emerging issues.
\end{abstract}

\section{Introduction}
\label{sec:1}
This paper comments on the recent approaches referred to as ``quantum-inspired'' in Information Retrieval (IR) research. IR is an area of classical computer science which covers problems related to the effective and efficient access to large amounts of stored information, where this information can be text documents, images, video, audio, etc~\citep{Baezayates_1999}. In ad-hoc information retrieval, which is the most common situation with which an IR system is concerned, input documents are transformed into a suitable representation for the IR system by an indexing process. A user expresses an information need in the form of a request that is formalised into a query by the IR system. The IR system compares the query against each document representation using a matching function determined by the adopted IR model. As a result of this comparison, the IR system produces a list of documents that are ranked from the most relevant to the least relevant, and this list is displayed to the user. This list reflects those documents that the IR system thinks are \emph{about} the query. In IR \emph{aboutness} is a semi-formalised notion which ``arises from an attempt to reason abstractly about the properties of documents and queries in terms of index terms''~\cite[p.19]{Rijsbergen_gir}. After examining the retrieved list, the user may provide feedback to the system by distinguishing documents that are relevant or modifying his/her initial information need. That is, IR is an interactive process --- the behaviour of the user modifies the system.

For several years the IR community has been considering some of the techniques and mathematical structures that have been used in quantum theory to model their tasks at hand.
These mathematical structures have typically been used by analogy, however, some of these approaches generate confusion. At least two papers have been written, targeting the IR community, to highlight some of the concerns about this program of research~\citep{Kantor_2007,Rieffel_2007}. However, on the basis of our observations of both the IR and Quantum Information communities, those who have a deep understanding of the physics of quantum theory and those who know merely the mathematical framework of quantum theory, tend to develop very different understanding of what has and has not been done in quantum-inspired IR. This paper is an introductory attempt to clarify the emerging issues and the authors hope it can benefit both communities.

The paper is organised as follows. Section~\ref{sec:2} discusses what IR researchers expect from Quantum Theory.
In Section 3, we provide a warning about the difference between storing information in a classical computer and storing quantum information on a quantum computer. Sections~\ref{sec:4} and~\ref{sec:5} describe various fundamental and advanced elements of QM, comparing them with their classical counterparts and discussing various proposed IR equivalences. The final discussion section lays out a number of challenges for this program of research if it is going to be successfully pursued.

\section{What do IR researchers want from Quantum Theory?}
\label{sec:2}

In this section we present our understanding of what IR researchers want from QM.

IR arose as a field driven by pragmatic considerations. Unlike other areas of classical computer science, finding a fundamental axiomatic formulation of the field has proven tricky, perhaps because of the intrinsic role humans play in the process and because information has proven considerably easier to capture mathematically than ``meaning". Quantum-inspired IR (QIIR) is an attempt to lay such a foundation.
The starting point can be traced back to viewing the field by the founder of QIIR,~\citet{Keith_towardslogic}, as a form of inference (from a document to a query or from a query to a document).
That is, a document is relevant to a query if it logically implies the given query and a measure of uncertainty is associated with such an implication. The uncertainty of implication is used to model the uncertainty of relevance.

This unifying viewpoint was motivated from prior work re-expressing three standard retrieval models (Boolean, Probabilistic and Co-ordination) in terms of uncertain implication~\citep{Keith_nonclassic_1986}.
In this proposal, it was suggested that accepting retrieval as inference allows us to ``speak quite clearly about theories of IR''. Theories themselves were described as ``a language together with axioms and rules of inference'', which provide a logic to perform this implication, and a measure of uncertainty of this implication~\citep{Keith_towardslogic}.

\Citet{Rijsbergen_experiment, Rijsbergen_gir} proposed to use the mathematical language of quantum mechanics, to describe the objects and processes in IR. It was also suggested that ``quantum mechanics provides a ready-made interpretation of this language. It is as if in physics we have an example semantics for the language, and as such it will be used extensively to motivate a similar but different interpretation for IR"~\citep[preface]{Rijsbergen_gir}.

While these intuitions are, at first glance, quite reasonable, there are difficulties in particulars.
There is an indication of such from the outset~\citep[preface]{Rijsbergen_gir}: ``The important notions in quantum mechanics, state vector, observable, uncertainty, complementarity, superposition and compatibility readily translate into analogous notions in information retrieval, and hence the theorems of quantum theory become available as theorems in IR". However, when we encounter some of the proposed IR-equivalences in Sections~\ref{sec:4} and~\ref{sec:5}, we will see that the majority do not admit a concrete or clear analogy which leads us to one of the main points of the paper: One should be very cautious about using the theorems of quantum theory as theorems in IR.

In this suggested quantum representation for IR, objects are represented within a Hilbert space, and a user can interact with these objects through \emph{measurement}. Measurement here is defined as application of linear operators (observables) to objects~(See Section~\ref{subsec:observables}). The outcome of this measurement is considered to be potentially probabilistic.
One aim for using this mathematical language was ``to apply the quantum theoretic way of looking at measurement to the finding of relevance in IR''~\citep[p.17]{Rijsbergen_gir}.

The essence of this proposal is the claim that using the mathematics of Hilbert space, which is also used by quantum theory, one can find a way of looking at information retrieval problems while being able to address questions of logic, probability and vector spaces within one framework. 

Where does the logic of inference come from? The properties of observables on Hilbert space form a non-Boolean lattice, which induces a non-Boolean logic; accordingly, the outcome of applying observables to these objects is controlled by this logic~\citep{Keith_qlogic_1996}. The outcomes of measurement have associated with them a probability as specified by the Gleason's Theorem~(See~\citet{Rijsbergen_gir} for further details). Further motivations for suggesting such a program were to model the complex notion of relevance~\citep{Rijsbergen_gir}, to capture the interaction between users and a retrieval system particularly with respect to the fact that both user and the IR system are allowed to evolve~\citep{Piwowarski_etal_2009} and to capture the user's context to help users finding information~\citep{Melucci_08}.

In the rest of this paper we provide some perspective on the above, our first step will be to clarify the difference between classical and quantum information processing.

\section{Warning: QIIR involves no physical quantum information}
\label{sec:3}

A warning similar to what we address in this section was raised in~\citet{Rieffel_2007} wherein the author briefly addressed \emph{``what is and isn't quantum information processing"}. We believe that it is well worth elaborating on this in the context of QIIR.

Mathematically a quantum state is a normalized vector of complex entries. The quantum state of even a small number of particles is vastly too large to ``write down" as classical bits. For example, a generic state of 50 two-level quantum systems (photons, atoms, spins etc) is represented by $2^{50}$ complex numbers, which if stored using single precision floating-point numbers (32-bit) would require 8192TB of space (about $10 ^{12}$ pieces of paper!) to store as a vector on a classical computer. Observables or unitary transformations to act on the state require even more than this. If we do have a lot of memory available and we write down the state of a number of particles then we have represented the quantum state by classical information. For such a classical representation of a quantum state we can of course read out any element of the state vector we choose, something we shall see is very different to how real quantum states are accessible in the physical world.

In the real world we can in principle \emph{physically prepare} 50 or more particles in a quantum state, despite normally not being able to write this state down classically. The preparation is not performed by somehow coupling a 8192TB hard drive to the particles and somehow copying in the elements of the desired state onto the systems! Rather, instructions on how to physically prepare the desired quantum state must be written on (a small number of) pieces of paper and passed on to a PhD student, who implements them in a lab. The instructions are classical information, they concern how to arrange pieces of equipment, and since they are also classical and on a small number of pages they could not contain $2^n$ parameters for any reasonably large $n$. We say they contain polynomial in $n$ ($poly(n)$) parameters.\footnote{With a little thought this shows that arbitrary (i.e. generic) quantum states are never prepared in a lab!}

Now, once the student has followed the instructions and prepared the particles in some quantum state we cannot then interact with those systems and simply ``read out" any particular entry of the state vector we choose. If we could do so we would, in fact, have extraordinary computational power~(see Appendix~\ref{appendix-qi-proof} for further details). The problem is that as soon as we interact with the systems we obtain indeterministic data and moreover we destroy the quantum state of the systems in so doing. So this is a big difference between a classical representation of a quantum state (the main focus of IR) and a genuinely quantum representation of a quantum state (which is a bunch of particles prepared to actually be in that state).

In fact, even though we might know the instructions the student followed to prepare the state, in general we can\emph{not} efficiently (i.e. using an algorithm on a classical computer that requires only poly(n) space and time) calculate the probabilities of measurement observables on this state.

These points are subtle even to practitioners of quantum physics and so bear some elucidation. 

So one might hope that perhaps the subset of quantum states which \emph{are} prepared in physics labs might be a subset we can both efficiently represent on a classical computer (for instance by storing just the instructions we gave the student!) and then be able to compute (classically) the probabilities of measurement outcomes for a specified observable on such a state. The observables themselves need to be ones that can be efficiently written down and parsed into experimental instructions too of course. The whole field of quantum information and computation is premised on the supposition (for which there is extremely good evidence) that this is not the case. We really do need to go into the lab, prepare and store the state \emph{quantumly} in actual physical systems and then perform physical measurements on those systems obtaining probabilistic data to obtain the well-documented information processing advantages proffered by quantum information.

So, when one talks of using the quantum state to represent some element of IR (e.g. a document, a query etc) if one means representation on a classical computer one is talking about an even much smaller subset of quantum states than even those a physics student can prepare in a lab. And now, since the state is just a vector of complex numbers stored in memory, one is able to do things that would be considered ``magical" within the framework of generic quantum physics  - such as reading out specific entries of the vector. In some sense this leads to the first question a quantum physicist might raise to a practitioner of IR: Why restrict yourself to the same set of rules that nature restricts us by, once you have supposedly adopted the Hilbert space framework but are then representing everything classically?

It is an open question whether there is any advantage to taking some element of IR, turning it into a procedure for creating some actual quantum state or implementing some observable in a lab and thereby letting genuine quantum information processing take place - that is, whether using \emph{quantum computers} is useful. But this is not what QIIR, at least in its original formulation, is aimed at.

The upshot is hopefully clear: QIIR as it stands today should be seen as adopting structural elements of quantum theory as a formal framework. Whether this is justified or useful is what the remainder of this article is about.

In the next section we further clarify the points highlighted in Section~\ref{sec:2} by providing introductory definitions of some of the fundamental elements of QM and further discuss their various proposed IR equivalences.

\section{Structural elements of QM {and} various proposed IR equivalences}
\label{sec:4}

The aim of this section and the next section is to give a clear picture of the basic elements of QM and suggested IR analogous notions. We also compare these elements with their classical counterparts.
We do not provide judgement on whether these analogies are sensible and concrete or not. The authors by no means claim that the list is exhaustive; These are merely a selected subset of features sufficient for the purpose of this commentary.

\subsection{Systems and States}

The status of the quantum state in physics is contentious. However all interpretations of quantum mechanics agree that quantum states are inferential objects: they at least play a role of allowing us to predict
the probabilities of future experimental outcomes. A Bayesian view of quantum mechanics~\citep*{Fuchs_2001} suggests that this is all they are (or ever can be), but there are certainly opposing opinions\citep{Valentini_2009, Deutsch_1999}. 

In classical physics the line is clear: epistemic states (states of knowledge/information) are probability distributions over some space, the points of which - the `ontic' or real states - encode the physical properties of the system. For instance; a single particle's ontic states are points in phase space, namely a position and a momentum. However, it is plausible an agent does not know the exact position or the exact momentum of a particle and so assigns a probability distribution (known in physics as a Liouiville distribution) over the phase space based on their information. Note that it is not impossible to think of quantum states in a similar way, but it does lead to some unusual looking descriptions of `underlying reality' and so it is not adopted by most physicists~\citep{Harrigan_Terry_2007,Harrigan_Spekkens_2010}.

Given that quantum states are inferential objects, we point out at this juncture that we could use a complete list of all probabilities of all possible measurement outcomes as a description of the state of the quantum system - an infinite list - but because of relationships between those probabilities we find using a finite vector in Hilbert space allows just as good inferences in a more compressed form. Why are state vectors called inferential objects? This is because QM does not allow us to read out any specific entries of the state vector we choose, and accordingly we need to perform measurement to predict the probabilities of future experimental outcomes from this inferential object. 

We now turn to proposed analogous IR notions to quantum systems and quantum states of which several appear in IR literature. However, we refer to two of the most developed ones, namely, representing the state of a \emph{document} or the state of a \emph{query} as an inferential object.

In the first approach, a given document is regarded as a quantum system and the problem of representing or indexing this document is considered analogous to that of representing quantum states of a physical system~\citep{Alvaro_etal2008}. The state of a document is therefore considered the inferential object. In IR vocabulary, this inferential object ``encapsulates the answers to all possible queries"~\citep{Rijsbergen_gir}. In principle the state assigned to the document could just be represented as a list of the probabilities (likelihoods) associated to every query - ie the queries play the role of measurement. However a more compressed version is obviously desirable.

In the second approach, a user's information need is considered as a quantum system and the state of this information need - the query - is the inferential object~\citep{Piwowarski_etal_2009} analogous to the quantum state, which can be used to calculate the outcome probabilities for all possible different measurements (documents) that may be performed.

\subsection{Measurement Observables}
\label{subsec:observables}
Measurements are described mathematically in quantum theory via applying an observable, that is, a Hermitian (self-adjoint) operator, or more precisely the projectors onto the eigenvectors of the observable in question, onto the state of a given system. Both the probabilities of an outcome and the post measurement state form part of the formalism and are deducible given a specific state and observable pair.

Both classical epistemic and quantum states are disturbed by a measurement - they both `collapse'. In the classical case this is not mysterious - when one gains information one changes the probability one assigns to a physical situation instantly. In the quantum case whether this is mysterious or not depends on whether one takes a Bayesian view or one's interpretation of the quantum state, an issue we are avoiding~\citep{Fuchs_2001}. What is certainly clear is that the disturbance in the quantum case is more radical due to the possibility of incompatible measurements.

Incompatible measurement refers to observables which are not jointly measurable, i.e. can not be measured at the same time, and measurement of one of them causes a disturbance to one's ability to predict the outcome of the other. Classical measurements are always compatible, quantum measurements can be either compatible or incompatible.

Interestingly, incompatible measurement is not always specific to quantum theory. There exist \emph{toy theories}~\citep*{toy_2007,ERLM_2011} which have incompatible measurements. These theories generate features people think of as typically quantum but the fact is that they are a restricted version of a classical theory. Another example of classical systems with incompatible variables can be found in~\citet{kirkpatrick-2003}. We will elaborate more on incompatibility in Section~\ref{sec:Incompatibility}.

We now return back to the question of what do we need to infer from a given system's state in IR?
Several possibilities come to mind.
If the task is finding documents relevant to a query, and the state of documents are considered as the inferential objects, then it seems natural that what should replace the role of measurements on the quantum system is a query~\citep{Rijsbergen_gir}.

The user of the IR system will then be the person who chooses a measurement to perform - i.e. who
chooses one from a set of possible queries to make. Posing a query by the user is regarded as applying a measurement operator onto the state of a given system. We need to assume that the query applied to any document yields a real number from 0 to 1, a number that,
in the quantum case, is the probability of that outcome.
These numbers are considered as indicative of likelihoods of user satisfaction. Thus the simple binary query to a document ``I am interested in pictures of an elegant motorcycle, do you contain one?'' should yield a real number $p \in[0; 1]$ which is higher if the document contains a picture that
the user will likely find elegant.

From this perspective the person who mathematically requires the inferential object is not the user per se, rather it is the person who is programming the IR system - the system programmer wants to infer whether any given document has a high or low likelihood of satisfying the user, given a future query to be chosen by users. Note that in this scenario we have
separated the role of \emph{measurer} from that of \emph{inferrer}, which typically we do not bother
doing in QM, but which here seems necessary.

For a single-term query, an observable corresponds to a yes/no question. For multi-term queries, the entire query is considered to be an observable which can be reduced to a combination of yes/no questions using the spectral decomposition theorem~\citep{Rijsbergen_gir}.

In the second approach, queries are considered as states of a quantum system, and
then documents correspond to the role of measurement, e.g.~\citep*{Piwowarski_what_2010}.

\subsection{Dynamics}

First we look at a closed quantum system which does not interact with any observer. A given quantum system evolves in time in a manner that depends both on intrinsic (internal) features and extrinsic (things nearby) ones. In this case the system changes continuously and the evolution operator is represented by the Schr\"odinger equation. This time evolution operator satisfies a number of criteria, including being a linear unitary operation, deterministic and differentiable in time. In quantum dynamics we are not free to choose whatever rules we wish to represent this time evolution. This evolution is in marked contrast to the already mentioned fact that upon measurement a quantum system's state changes discontinuously - it collapses into one of the observable's eigenstates with some probability (itself dependent on the geometry of Hilbert space via Born's Rule).

As the evolution operator must map an allowed state of the system to another allowed state (most operations would take us to invalid states) there is therefore a trade-off between the allowed states of a theory and the allowed dynamics of the theory~\citep{Joh_barrett_2007}. Allowed state refers to those states that "correspond to physically possible states of a system"~\citep{Joh_barrett_2007}.
It is worth mentioning here that there exist theories that are neither quantum nor classical but their dynamic is classical~(See~\citet{Joh_barrett_2007} for some examples of these theories).

In the context of interactive IR, it is suggested that if a user's query is represented as a quantum state then this state evolves when the user interacts with the system. Two different dynamics are suggested~\citep*{Piwowarski_etal_2009}. Firstly, during the interaction of the user with the IR system, the system's understanding of user's information need changes. Secondly, a user may change his/her information need after receiving some information during the interactions with the IR system. Any interaction between the IR system and the user is modelled as a measurement, and measurement is modelled using projection operators (yes/no observable), hence following the discontinuous type of dynamics as mentioned above.

In the second approach, if a document is represented as a quantum state then this state may need to be evolved discontinuously~(by the inferrer) when the user provides feedback on the relevance of the given document\citep{Sachi_thesis08}.

\section{Quantum Phenomena}
\label{sec:5}

In this section we survey some of the more controversial quantum mechanical concepts which induce confusion within the quantum-inspired IR approaches.
We take a closer look at the IR-theoretic analogue(s) of the notions of randomness and incompleteness, incompatibility, statistical mixture and superposition, non-orthogonality, and quantum interference. We end this section by representing some issues with the use of quantum logic for IR.

\subsection{Randomness and Incompleteness}

In this section we compare the notion of classical and quantum randomness, and highlight few questions regarding the application of quantum viewpoint of information processing in IR. The intention of authors here is neither to define any quantitative measure of randomness, nor to defend the usefulness of randomness as a resource in IR\footnote{For an interesting example of using randomness as a resource in designing rich user experiences for listening to music see~\citet*{Leong_etal2006}.}.

Classically the randomness (unpredictability) of the measurement outcome is viewed as observer's ignorance about some pre-determined properties, i.e. it reflects the incomplete knowledge of the observer about the system. This knowledge can be improved or even completed via further measurements and the acquisition of more information.

In the quantum case, the outcome of a specific quantum measurement is unpredictable, however this unpredictability may or may-not be considered to reflect a lack of knowledge about some hidden variables according to one's interpretational leanings. Within the framework of pure quantum theory we can say only that given a pure quantum state, predictability cannot be improved. But of course the predictability is only perfect for a small set of the possible measurements, for most of them the outcomes are probabilistic. And attempts to further refine that information - as can be done classically - run into the problem of incompatibility. There is some maximum amount of information that we can have about a quantum system, but this does not give us the complete information that would be necessary to predict the outcome of \emph{any} measurement: \emph{maximum information about a quantum system as represented by a pure quantum state vector is not necessarily complete information}~\citep{Fuchs_2001,Fuchs_etal_1996,toy_2007}.

To understand the nature of unpredictability in IR, we are required to answer the following question:
Is a user's searching for information as intrinsically unpredictable as that encountered in QM or is IR randomness simply related to the classical concept of incomplete knowledge? In other words, whether we have a more concrete analogy between IR and an incomplete knowledge classical system as opposed to a quantum system.
The answer will depend in some circumstances on whether there is a natural notion of incompatible measurements in IR (see next section). If the answer is positive, then IR notion of randomness is a sensible analogous notion to quantum randomness. Otherwise, we are associating an inherent randomness to a system, only because the system is complex and we do not have complete knowledge about it.\footnote{A concrete question (partly addressed in~\citet{PavlovicD_QI08}) is whether quantum randomness can improve the task of modelling information retrieval systems.}

In QIIR the probabilities of quantum theory are generically carried over to relate in some manner to `document rankings' - to the likelihood of user satisfaction with a document\footnote{The utilised probability notion should not be used as an objective probability, i.e. it is not that we expect if one repeat an experiment multiple times then a certain fraction of them would result in satisfaction.}.
This inferential uncertainty is basically uncertainty about a user's preferences, goals and judgements.
In fact, several studies showed that users' relevance criteria fluctuate depending on their interests, intentions and goals~\citep*{Law_etal_2006}. This implies that each time a user judges whether a document is relevant to a given query, he or she may provide a different judgement depending on the relevance criteria at the time. Does it follow that the user's behaviour is completely unpredictable and we cannot know the user at all?
As the users of an IR system are diverse in goals, intentions, and many other factors, one research question to investigate here is to what extent we can acquire the maximal information about user(s) of an IR system, of course considering the variety of information retrieval tasks. 
In this sense we are treating the users as a system - there is a sample space of all possible users. If a (classical or quantum) state is meant to capture our uncertainty about the outcomes of future experiments then from this perspective one can consider associating states to users and not to queries (or documents).

\subsection{Incompatibility}
\label{sec:Incompatibility}

The notion of \emph{incompatible measurement} refers to observables which are not simultaneously measurable, measuring one of them causes a random disturbance in the other -  so different outcomes will be obtained depending on which is measured first. Quantum incompatibility is manifested mathematically by the non-commutativity of the observables, which means the order of successive measurements affects the outcome of the measurements.

Before referring to possible benefits of using incompatibility in IR, it should be noticed that the notion of incompatibility does not exclusively belong to quantum theory. An example of a classical system with incompatible variables is already addressed in~\citet{kirkpatrick-2003}, and example of classical non-commutativity of measurements can be found in the Toy Theory of Robert W.~Spekkens~\citeyearpar{toy_2007} and classical Liouville mechanics with an epistemic restriction~\citep{ERLM_2011}.

If the state (i.e. inferential object) refers to a document and a query to measurement, an intuitive use of incompatibility in IR is where a document passes through a sequence of queries in which reversing the order does not lead to the same results.
A supporting evidence for the existence of such queries refers to a study~\citep*{Busemeyer_etal_2011} in which a story (i.e. a document) is represented to a user and then successive queries are answered by the user about the likelihood of certain statements about the story. The order of these questions has been shown to be significant.
An example suggested in~\citet{Rijsbergen_gir} refers to treating the relation between \emph{aboutness} and \emph{relevance} of a query as incompatibility between observables. It has been argued that incompatibility between predicates happens because of interaction, e.g. if one asks an observer questions regarding \emph{aboutness} and sequentially a question about its \emph{relevance}, by repeating the first question one might get a different answer than when it was originally asked.

If the state refers to a query and a document to measurement, one suggestion for using incompatibility in IR is where a user is provided with two different documents, $d_1$ and $d_2$. The user may first study $d_1$ and then $d_2$ and if we again ask user's opinion about $d_1$, the opinion may have been changed. This implies that the order of documents matters and the `observables' correspond to reading each of these documents do not always commute, e.g.~\citep{Piwowarski_etal_2009}. A supportive evidence for the influence of presentation order on user judgments can be found in~\citet*{Eisenberg_etal1988}.

One fundamental question to investigate would be whether we have enough examples of incompatibility in IR and if we can provide a strong case to justify this analogy. As classical logic is all about compatible observables, finding appropriate answer to the above research question will have a great impact on another question which will come up about incompatible observables and that is to what extent we need to use \emph{quantum logic} for IR.

\subsection{Statistical Mixture and Superposition}
\label{sec:superposition}
Mixed and superposed quantum states are two of the other notions from quantum theory that are suggested to have IR analogous notions. First let recall what is a mixed state in Quantum Mechanics.

\subsubsection{Mixed states}

Assume that the only information an observer knows about a given system is that the system is in state $A$ with probability $P_A$ and in state $B$ with probability $P_B$. This uncertain information about the state of the system can be represented by a mixed state. A mixed state is represented by a density matrix which is a mathematical object that allows to represent experimental scenarios involving a randomization in choice of measurement, preparation of states, etc.
A mixed state corresponds to a classical probability distribution over a set of \emph{pure} quantum states in which
a pure states refer to a state that can not be represented as a mixture of other quantum states.

Until one is clear about the role of pure states within QIIR it is somewhat difficult to be sure about the role of mixed states.
An intuitive suggestion to use the concept of mixed states in IR is to represent an inferential object with some inherent uncertainty. One scenario would be the representation of a single-term query that involve a randomization in choice of query terms. This scenario refers to the case where the user's query is a single-term query but the inferrer is not aware of  the exact query term that the user is asking. As an example consider a user who randomly chooses query term $t_1$ with probability $p_1$, query term $t_2$ with probability $t_2$, and $t_3$ with probability $p_3$ as his/her information need. In this scenario the inferrer who is not aware of the exact query term that the user is asking, uses a mixed state to represent the user's query. As a result of this representation, the probability that the inferrer will measure for each document would be the convex combination of the three cases. This scenario is close to what mixed states are used for in QM and can be extended to mixtures of superposed quantum states as addressed in~\citet{Piwowarski_what_2010}.

Another suggestion in IR is to use the concept of mixed states to represent a set of inferential objects, each with its own importance degree~\citep{Rijsbergen_gir}. One such scenario would be the representation of multi-term queries, i.e. combinations of query terms with weight coefficients. An example is where the user's query contains three query terms ``$t_1\ t_2\ t_3$'', associated with importance degree of $p_1, p_2, p_3$, respectively. According to the suggested approach, the inferrer assigns a mixture representation to this query. However, any attempt to interpret this representation under the light of the mixed state used in QM would mean that the system (i.e. query) is just in state $t_1$ with probability $p_1$, in state $t_2$ with probability $p_2$ and in state $t_3$ with probability $p_3$. This interpretation differs from the expectation in IR that the user's query is meant to address all three query terms at once.
Accordingly this suggestion reflects a fundamental difference with the concept of mixed states in QM.
Other examples of exploiting this scenario in IR include representing documents in a cluster~\citep{Rijsbergen_gir} and representing the set of documents that user judges as relevant~\citep{Rijsbergen_gir}.

The message to be noticed here is that mixed states are merely a classical distributions over a set of quantum states, and if the goal is to adopt the quantum viewpoint of information processing, a concrete analogy with the concept of mixed states in QM is expected. One question to investigate here is to what extent the uncertainty inherent in density operators matches with the type of mixture we need in IR.

\subsubsection{Superposition}

In quantum theory one can create superposition states of a single quantum system. For example a given system can be in a superposed state $|\phi\rangle$ = $\alpha |A\rangle$ + $\beta |B\rangle$, where
$|.\rangle$ denotes a state vector in the Hilbert space and refers to the Dirac notation used in QM, $\alpha^{*}\alpha$ + $\beta^{*}\beta$ = 1, and $^{*}$ denotes complex conjugate.

It is crucial to note that the system in a superposed state is in a genuinely new state. It is fundamentally different from a system that is in state $|A\rangle$ or one in state $|B\rangle$. But it is also fundamentally different from a system that is just `in state $|A\rangle$ with probability $p$ and in state $|B\rangle$ with probability $1-p$'. That situation is the case of a mixed state, referred to above. Superposition is an operation that takes in two (or more) states and combines them into a completely different state within the Hilbert space. Inferences from this new state differ from those we make from $|A\rangle$ and $|B\rangle$. A superposition of two (or more) possibilities is not the \emph{conjunction (and)} of those possibilities nor the \emph{disjunction (or)} of those possibilities. It should also be pointed out that the same state can be written as a coherent superposition of \emph{many different} pairs of states.

In IR, there are two different viewpoints towards the significance of the concept of superposition. In a first view, van Rijsbergen mentioned that ``it is not clear whether the difference between a superposition and a mixture of states plays a significant role in IR"~\citep[p.95]{Rijsbergen_gir}.
Given that van Rijsbergen also mentioned~\citep[preface]{Rijsbergen_gir} that the important quantum phenomena including superposition readily translates into analogous notions in IR,
this viewpoint is a somehow contradictory view since it is the difference between superposition and mixture that differentiates classical from quantum physics. Several other IR researchers do, however, consider the concept of superposition a significant concept for IR.
There are several attempts to provide analogous notions to superposition, but these approaches generally address a concept which is very different than quantum superposition.

One such attempt is to use the notion of superposition in query representation~\citep{Piwowarski_what_2010}. An example of this attempt is when the query ``Pizza delivered in Cambridge (UK)'' would be represented by a (normalised) linear combination (or, as it was claimed, `superposition') of the state vectors associated with ``I want a pizza'', and the one associated to ``I want it to be delivered in Cambridge (UK)''. As we addressed before, a weighted coherent superposition of two possibilities is not the \emph{and} of those possibilities nor the \emph{or} of those possibilities. Accordingly in this example, the advantage of calling this combination as `superposition` is not clear.
Other suggested analogue(s) of the notion of superposition can be found in~\citet*{Melucci_Keith_2011}.

A final point to note about superpositions is the following. If quantum states $|A\rangle$ and $|B\rangle$ are orthogonal, the state $\alpha |A\rangle$ + $\beta |B\rangle$ is not orthogonal to either. Is there a QIIR analogue of non-orthgonality to help us identify a QIIR analogue of superposition? If we are associating states to documents then two such states would be orthogonal if there is a query that yields a relevance of 1 on the one document and 0 on the other. Conversely, if we are associating states to queries then orthogonal queries are ones for which there is some document that is completely relevant for one and irrelevant for the other. The non-orthogonal combination of the documents or queries should be something that yields some relevance in-between 0 and 1 for such cases. It seems plausible that QIIR examples of such phenomena could be found.

There is a catch however. Non-orthonality arises even in classical probability theory. It arises for overlapping distributions. That is, there are points in the sample space which can be associated with more than one probability distribution. Overlapping probability distributions share many features of quantum non-orthogonality. They cannot be distinguished perfectly, they cannot be cloned~(identically copied) and so on. The toy theories mentioned above~\citep{toy_2007,ERLM_2011} all make extensive use of this feature. Determining what is genuinely quantum about non-orthogonality is subtle -
if one claims that there would be some quantitative benefits for using quantum states in IR, first should show that the same outcomes can not be achieved using classical non-orthogonality.

To conclude, in order to investigate whether the difference between a superposition and a mixture of states plays a significant role in IR we suggest that there should be a clear concept of interference - the fundamental phenomenon that is a consequence of this difference.

\subsection{Quantum Interference}
\label{interference}
Quantum interference is a phenomenon in which the probability of two mutually exclusive events $A$ and $B$, is not $P_{A}+P_{B}$ (the sum of their individual probabilities) but is rather $P_{A}+P_{B}+f(P_{A},P_{B},\theta)$ where $f$ is some function of $P_{A}, P_{B}$, and some set of other parameters (denoted $\theta$) that are either part of the quantum state or are part of the measurement. A simple interference commonly encountered has
$f(P_{A},P_{B},\theta)=2\sqrt P_{A}\cdot\sqrt P_{B}\cdot cos\theta$.

To what does quantum interference correspond in IR? Let us focus on the case where we are assigning states to documents for concreteness. In some sense it should be a process of document
merging which says something like: Combining two possible inferential objects associated with a document yields  an inferential object such that new queries have
outcomes that are \emph{not} formed by probabilistic mixture of the likelihoods for the previous
inferential objects. That is, sometimes the best new inferential object lies
\emph{in-between} the original two in a \emph{non-classical} manner.

Here is an example of our own devising. Consider some ancient manuscript, such as the Bible, for which the
oldest known pieces are fragmentary. D1 could be one such piece, and D2 another. The `real' document - the system - is the original Bible, based on D1 and D2 we might assign different indexes and inferential objects - `states' - according to which we answer questions about whether the bible will satisfy some need of the user. Typically
we find that D1 and D2 agree on the text in many places, however both have text which
is not contained in the other. Moreover, in places where they disagree, scholarly input is
required to analyse the context, to look for possible scribal errors and so on. Therefore the
final reconstructed document D3 is close to D1 and D2, but it is not a simple merging of
them.

Consider now someone making a series of queries on the documents. It is perfectly
possible that the likelihood they assign to D3 satisfying the users information need is not simply a simple probabilistic mixture
of the likelihood for D1 and D2 individually - it is not as if the scholar who produced D3
did so by, wherever there was discrepancy, choosing D1 with probability 2/3 and D2 with
probability 1/3 say. Therefore we have a concrete situation where the inferential object we
would associate with D3 may well be ``non-orthogonal" (in the operational quantum sense)
to that associated with D1 and D2. Of course the functional form of the interference is unlikely
to follow that of QM!.

Other proposed analogous notions to interference could be along the lines that the information in two
documents conflicts (or corroborates), so returning both of them might result in lower (or
higher) user satisfaction than returning each of them individually~\citep{Guido_ecir2010}.
Similarly, the information in two topics, in a long document or document containing multiple topics or subtopics, may conflict~\citep*{Wang_etal_2010}. Similar analogies can be formed on the query side as well, e.g. merging two query terms together~\citep{Piwowarski_etal_2009}.

In quantum mechanics there is a specific functional form that interference takes, due to the geometry of Hilbert space. If we wish to push for a specific type of interference such as those above then the next step is to justify its specific mathematical structure.
There are many general probabilistic theories manifesting interference phenomena in different forms - even within physics it is hard to justify why nature took the specific functional form it did. It therefore seems extremely unlikely that interference in QIIR can easily justify it; if it did we would surely gain new knowledge about quantum physics! The discussion section details about using alternative theories to model interference and other quantum effects.

\subsection{A framework to combine Probability, Logic and Vector Spaces?}
\label{sec:logic}

One of the main motivations for suggesting to use quantum theory in IR was using its underlying mathematical language as a framework to combine probability, logic and vector spaces in one formalism~\citep{Rijsbergen_gir}. The suggestion refers to the fact that Quantum Theory provides this framework via its standard version of quantum logic and via Gleason's theorem.

The standard version of quantum logic refers to the set of all closed subspaces of a Hilbert space or the set of all orthogonal projection operators~(the 2-valued observables) which constitute an orthomodular lattice which is not distributive. This orthomodular lattice determines a sort of logic and a probability measure can be specified on this lattice of subspaces using the Gleason's theorem.

Many researchers refer to this version as a failed program of research. It is certainly no longer heavily pursued in physics. For a list of some of these reasons
see~\citet{Coecke_proposal_qlogicfail}.

One particularly relevant issue, addressed first by Foulis and Randall~\citeyearpar{Randall_1979}, relates to describing a composite system via a tensor product. A composite quantum system involves more than one subsystem, wherein the Hilbert space for this system is represented as the tensor product of the Hilbert spaces for each subsystem. Foulis and Randall showed that it is not possible to formulate a tensor product for orthomodular lattices. This means that the tensor product of orthomodular lattices is not necessarily an orthomodular lattice.

It would seem inevitable that for something to be useful as an axiomatic foundation for IR that it be able to deal with the case of multiple systems (regardless of whether `the system' refers to documents, queries or something else). As a natural consequence of the failure of standard quantum logic for composite systems, some of the suggestions for using tensor products in IR, e.g. combining different representations of a given document by means of tensor products and using the notion of non-separable states to formalise the complex interdependent relationships among the different representations~\citep*{Ingo-Polyrepresentation} are not necessarily compatible with some of the underlying motivations for the program.

More general mathematical structures developed later in the area of quantum logic in search for a genuine version of quantum logic for quantum theory, including orthoalgebras (a generalization of orthomodular lattices), effect algebras (a generalization of orthoalgebra) and category theoretic structures~\citep{BobCoecke_picture,Bobcoecke_maincategory}\footnote{For an application of the categorical theory version of quantum logic on the domain of natural language processing see~\citet*{ClarkCoeckeSadr}.}. For a review and a list of references showing the development of quantum logic, including orthoalgebras and effect algebras consult~\citet*{Bob_2000}. The potential benefits of these alternative logics for IR remain to be seen.

\section{Discussion and suggestions for future research}
\label{sec:6}

In this paper we surveyed some of the elements of QM and various proposed IR equivalences.
The main weakness of such proposals as discussed in Sections~\ref{sec:4} and~\ref{sec:5} is that they do not have a convincingly concrete analogy with elements of QM.
This implies a challenge for this program of research to identify what are the requirements for IR to have a concrete analogy with QM, and to investigate whether assuming those requirements for IR problems are reasonable.
Even if we come up with a concrete analogy between IR and quantum phenomena, e.g. strong cases for the existence of incompatibility in measurement, sufficiently similar types of randomness, etc, and we are looking for a theory to reproduce those phenomena, then it should be noted that the mathematics of Hilbert space and its induced logic is merely a tool to perform further foundational research in IR, and one may always expect that a more appropriate tool could arise.
Van Rijsbergen in a keynote talk at SIGIR 2006~\citep{Rijsbergen_keynote_sigir} addressed that the elements of logic in this tool can be challenged. He mentioned that ``The assumption that closed linear subspaces will be the elements of our logic can be challenged, as perhaps a construction with different elements is possible". One can consult Section~\ref{sec:logic} for examples of some of these alternative tools, wherein we referred to some of the existing replacements for the standard quantum logic.
We would like to point out that the probabilistic framework in this tool can also be challenged. There are many general probabilistic theories manifesting features very similar to quantum phenomena that can possibly also be used to help formalise the foundations of IR.

We now briefly introduce some of the possible alternatives for the probabilistic framework. As finding the simplest theory for IR uses would be desirable, being aware of such theories is useful. Although in physics none of these theories have been explored as rigorously as QM, nothing stops IR researchers from exploring the options to find the most appropriate solution for IR.

The first category of theories refers to \emph{In-Between Theories} which are theories that are closer to classical theory as opposed to quantum theory, and have no connections with physics but reproduce a great number of quantum phenomena. As examples we refer to the Toy Theory of~\citet{toy_2007} and classical Liouville
mechanics with an epistemic restriction~\citep{ERLM_2011}. These are both theories built on top of regular probability theory but with a single restriction that forbids someone ever being 100 percent sure of the true value of the random variable.
In these theories, a simple restriction on the available knowledge for the observer allows one to model a very large number of features commonly thought to be quantum phenomena. Examples of these features include non-commutativity of measurements, a form of interference, superposition, the multiplicity of convex decompositions of a mixed state, mutually unbiased bases, the EPR paradox, teleportation and many others.

The second category of alternative theories refers to \emph{Generalised Theories}. This category of theories is a broader range of theories than classical and quantum, i.e.~is
sufficiently general to include quantum theory, classical probabilistic theories and many other theories.
One example of this category refers to theories that are classified under the category of \emph{Generalised Probabilistic Theories Framework}~\citep{Joh_barrett_2007} which is a conservative generalisation of classical probability theory. The fundamental elements we listed in Section~\ref{sec:4} are those structural features that all theories within this framework have in common (i.e. states, measurement observables, dynamics). In fact certain quantum phenomena appear in all of these theories except classical theories!
Therefore the onus is on IR researchers to justify singling out quantum theory from the others, since at present the arguments in favour of quantum inspired IR merely look at qualitative analogies with these phenomena.

Alternatively, it is possible that the Generalised Probabilistic Theories Framework is still not general enough for the tasks in IR and one may need to write down a mathematical framework that generates the specific required features, using e.g.~\emph{Convex Operational Theories}~\citep{Barnum_etal_2011} or \emph{Category Theoretical Structures}~\citep{Bobcoecke_maincategory,BobCoecke_picture}, all of which contain quantum theory as a special case.

The above is not claimed to be an exhaustive list, but are merely examples to make the reader aware of other possibilities.
In our opinion whether there are actually any practical benefits of using a specifically quantum inspired model remains an open and challenging topic for the future of IR theory.

\section{Acknowledgements}
The authors gratefully acknowledge S.Arafat, M. Leifer and B. Coecke for discussions on these issues.
This research was supported by the Imperial College EPSRC Strategic Fund.

\bibliographystyle{apacite}
\bibliography{CommentaryQIIR}

\begin{thebibliography}{}

\bibitem [\protect \citeauthoryear {%
Abramsky%
\ \BBA {} Coecke%
}{%
Abramsky%
\ \BBA {} Coecke%
}{%
{\protect \APACyear {2004}}%
}]{%
Bobcoecke_maincategory}
\APACinsertmetastar {%
Bobcoecke_maincategory}%
\begin{APACrefauthors}%
Abramsky, S.%
\BCBT {}\ \BBA {} Coecke, B.%
\end{APACrefauthors}%
\unskip\
\newblock
\APACrefYearMonthDay{2004}{}{}.
\newblock
{\BBOQ}\APACrefatitle {A categorical semantics of quantum protocols} {A
  categorical semantics of quantum protocols}.{\BBCQ}
\newblock
\BIn{} \APACrefbtitle {{Proceedings of the 19th {IEEE} conference on Logic in
  Computer Science ({LiCS'04})}} {{Proceedings of the 19th {IEEE} conference on
  Logic in Computer Science ({LiCS'04})}}\ (\BPGS\ 415--425).
\newblock
\APACaddressPublisher{}{IEEE Computer Science Press}.
\PrintBackRefs{\CurrentBib}

\bibitem [\protect \citeauthoryear {%
Arafat%
}{%
Arafat%
}{%
{\protect \APACyear {2008}}%
}]{%
Sachi_thesis08}
\APACinsertmetastar {%
Sachi_thesis08}%
\begin{APACrefauthors}%
Arafat, S.%
\end{APACrefauthors}%
\unskip\
\newblock
\APACrefYear{2008}.
\unskip\
\newblock
\APACrefbtitle {Foundations research in information retrieval inspired by
  quantum theory} {Foundations research in information retrieval inspired by
  quantum theory}\ \APACtypeAddressSchool {Ph{D} Thesis}{}{}.
\unskip\
\newblock
\APACaddressSchool {}{University of Glasgow, UK}.
\PrintBackRefs{\CurrentBib}

\bibitem [\protect \citeauthoryear {%
Baeza-Yates%
\ \BBA {} Ribeiro-Neto%
}{%
Baeza-Yates%
\ \BBA {} Ribeiro-Neto%
}{%
{\protect \APACyear {1999}}%
}]{%
Baezayates_1999}
\APACinsertmetastar {%
Baezayates_1999}%
\begin{APACrefauthors}%
Baeza-Yates, R.%
\BCBT {}\ \BBA {} Ribeiro-Neto, B.%
\end{APACrefauthors}%
\unskip\
\newblock
\APACrefYear{1999}.
\newblock
\APACrefbtitle {Modern Information Retrieval} {Modern information retrieval}.
\newblock
\APACaddressPublisher{}{Addison Wesley}.
\PrintBackRefs{\CurrentBib}

\bibitem [\protect \citeauthoryear {%
Barnum%
\ \BBA {} Wilce%
}{%
Barnum%
\ \BBA {} Wilce%
}{%
{\protect \APACyear {2011}}%
}]{%
Barnum_etal_2011}
\APACinsertmetastar {%
Barnum_etal_2011}%
\begin{APACrefauthors}%
Barnum, H.%
\BCBT {}\ \BBA {} Wilce, A.%
\end{APACrefauthors}%
\unskip\
\newblock
\APACrefYearMonthDay{2011}{}{}.
\newblock
{\BBOQ}\APACrefatitle {Information Processing in Convex Operational Theories}
  {Information processing in convex operational theories}.{\BBCQ}
\newblock
\APACjournalVolNumPages{Electronic Notes in Theoretical Computer
  Science}{270}{}{3--15}.
\PrintBackRefs{\CurrentBib}

\bibitem [\protect \citeauthoryear {%
Barrett%
}{%
Barrett%
}{%
{\protect \APACyear {2007}}%
}]{%
Joh_barrett_2007}
\APACinsertmetastar {%
Joh_barrett_2007}%
\begin{APACrefauthors}%
Barrett, J.%
\end{APACrefauthors}%
\unskip\
\newblock
\APACrefYearMonthDay{2007}{}{}.
\newblock
{\BBOQ}\APACrefatitle {{Information processing in generalized probabilistic
  theories}} {{Information processing in generalized probabilistic
  theories}}.{\BBCQ}
\newblock
\APACjournalVolNumPages{Physical Review A}{75}{3}{032304}.
\PrintBackRefs{\CurrentBib}

\bibitem [\protect \citeauthoryear {%
Bartlett%
, Rudolph%
\BCBL {}\ \BBA {} Spekkens%
}{%
Bartlett%
\ \protect \BOthers {.}}{%
{\protect \APACyear {2011}}%
}]{%
ERLM_2011}
\APACinsertmetastar {%
ERLM_2011}%
\begin{APACrefauthors}%
Bartlett, S\BPBI D.%
, Rudolph, T.%
\BCBL {}\ \BBA {} Spekkens, R\BPBI W.%
\end{APACrefauthors}%
\unskip\
\newblock
\APACrefYearMonthDay{2011}{}{}.
\newblock
\APACrefbtitle {{Reconstruction of Gaussian quantum mechanics from Liouville
  mechanics with an epistemic restriction}.} {{Reconstruction of Gaussian
  quantum mechanics from Liouville mechanics with an epistemic restriction}.}
\newblock
\APAChowpublished {arXiv:1111.5057v1}.
\PrintBackRefs{\CurrentBib}

\bibitem [\protect \citeauthoryear {%
Busemeyer%
, Pothos%
, Franco%
\BCBL {}\ \BBA {} Trueblood%
}{%
Busemeyer%
\ \protect \BOthers {.}}{%
{\protect \APACyear {2011}}%
}]{%
Busemeyer_etal_2011}
\APACinsertmetastar {%
Busemeyer_etal_2011}%
\begin{APACrefauthors}%
Busemeyer, J\BPBI R.%
, Pothos, E\BPBI M.%
, Franco, R.%
\BCBL {}\ \BBA {} Trueblood, J.%
\end{APACrefauthors}%
\unskip\
\newblock
\APACrefYearMonthDay{2011}{}{}.
\newblock
{\BBOQ}\APACrefatitle {A quantum theoretical explanation for probability
  judgment errors} {A quantum theoretical explanation for probability judgment
  errors}.{\BBCQ}
\newblock
\APACjournalVolNumPages{Psychological Review}{108}{}{193-218}.
\PrintBackRefs{\CurrentBib}

\bibitem [\protect \citeauthoryear {%
Caves%
\ \BBA {} Fuchs%
}{%
Caves%
\ \BBA {} Fuchs%
}{%
{\protect \APACyear {1996}}%
}]{%
Fuchs_etal_1996}
\APACinsertmetastar {%
Fuchs_etal_1996}%
\begin{APACrefauthors}%
Caves, C\BPBI M.%
\BCBT {}\ \BBA {} Fuchs, C\BPBI A.%
\end{APACrefauthors}%
\unskip\
\newblock
\APACrefYearMonthDay{1996}{}{}.
\newblock
{\BBOQ}\APACrefatitle {Quantum Information: How Much Information in a State
  Vector?} {Quantum information: How much information in a state
  vector?}{\BBCQ}
\newblock
\BIn{} A.~Mann\ \BBA {} M.Revzen\ (\BEDS), \APACrefbtitle {The Dilemma of
  {Einstein}, {Podolsky} and {Rosen}, 60 Years Later} {The dilemma of
  {Einstein}, {Podolsky} and {Rosen}, 60 years later}\ (\BPGS\ 226--257).
\PrintBackRefs{\CurrentBib}

\bibitem [\protect \citeauthoryear {%
Caves%
, Fuchs%
\BCBL {}\ \BBA {} Schack%
}{%
Caves%
\ \protect \BOthers {.}}{%
{\protect \APACyear {2002}}%
}]{%
Fuchs_2001}
\APACinsertmetastar {%
Fuchs_2001}%
\begin{APACrefauthors}%
Caves, C\BPBI M.%
, Fuchs, C\BPBI A.%
\BCBL {}\ \BBA {} Schack, R.%
\end{APACrefauthors}%
\unskip\
\newblock
\APACrefYearMonthDay{2002}{}{}.
\newblock
{\BBOQ}\APACrefatitle {{Quantum probabilities as Bayesian probabilities}}
  {{Quantum probabilities as Bayesian probabilities}}.{\BBCQ}
\newblock
\APACjournalVolNumPages{Physical Review A}{65}{2}{022305}.
\PrintBackRefs{\CurrentBib}

\bibitem [\protect \citeauthoryear {%
Coecke%
}{%
Coecke%
}{%
{\protect \APACyear {2008}}%
}]{%
Coecke_proposal_qlogicfail}
\APACinsertmetastar {%
Coecke_proposal_qlogicfail}%
\begin{APACrefauthors}%
Coecke, B.%
\end{APACrefauthors}%
\unskip\
\newblock
\APACrefYearMonthDay{2008}{}{}.
\newblock
\APACrefbtitle {{The Road to a New Quantum Formalism - Categories as a Canvas
  for Quantum Foundations}.} {{The Road to a New Quantum Formalism - Categories
  as a Canvas for Quantum Foundations}.}
\newblock
\APAChowpublished
  {{http://www.cs.ox.ac.uk/people/bob.coecke/FQXi.pdf~({S}ection~1)}}.
\PrintBackRefs{\CurrentBib}

\bibitem [\protect \citeauthoryear {%
Coecke%
}{%
Coecke%
}{%
{\protect \APACyear {2009}}%
}]{%
BobCoecke_picture}
\APACinsertmetastar {%
BobCoecke_picture}%
\begin{APACrefauthors}%
Coecke, B.%
\end{APACrefauthors}%
\unskip\
\newblock
\APACrefYearMonthDay{2009}{}{}.
\newblock
{\BBOQ}\APACrefatitle {Quantum Picturalism} {Quantum picturalism}.{\BBCQ}
\newblock
\APACjournalVolNumPages{Contemporary Physics}{51}{}{59-83}.
\PrintBackRefs{\CurrentBib}

\bibitem [\protect \citeauthoryear {%
Coecke%
, Moore%
\BCBL {}\ \BBA {} Wilce%
}{%
Coecke%
\ \protect \BOthers {.}}{%
{\protect \APACyear {2000}}%
}]{%
Bob_2000}
\APACinsertmetastar {%
Bob_2000}%
\begin{APACrefauthors}%
Coecke, B.%
, Moore, D.%
\BCBL {}\ \BBA {} Wilce, A.%
\end{APACrefauthors}%
\unskip\
\newblock
\APACrefYearMonthDay{2000}{}{}.
\newblock
{\BBOQ}\APACrefatitle {Operational quantum logic: an overview} {Operational
  quantum logic: an overview}.{\BBCQ}
\newblock
\BIn{} B.~Coecke, D.~Moore\BCBL {}\ \BBA {} A.~Wilce\ (\BEDS), \APACrefbtitle
  {{Current Research in Operational Quantum Logic: Algebras, Categories and
  Languages, Fundamental Theories of Physics series}} {{Current Research in
  Operational Quantum Logic: Algebras, Categories and Languages, Fundamental
  Theories of Physics series}}\ (\BPGS\ 1--36).
\newblock
\APACaddressPublisher{}{Kluwer Academic Publishers}.
\PrintBackRefs{\CurrentBib}

\bibitem [\protect \citeauthoryear {%
Coecke%
, Sadrzadeh%
\BCBL {}\ \BBA {} Clark%
}{%
Coecke%
\ \protect \BOthers {.}}{%
{\protect \APACyear {2010}}%
}]{%
ClarkCoeckeSadr}
\APACinsertmetastar {%
ClarkCoeckeSadr}%
\begin{APACrefauthors}%
Coecke, B.%
, Sadrzadeh, M.%
\BCBL {}\ \BBA {} Clark, S.%
\end{APACrefauthors}%
\unskip\
\newblock
\APACrefYearMonthDay{2010}{}{}.
\newblock
{\BBOQ}\APACrefatitle {Mathematical Foundations for a Compositional Distributed
  Model of Meaning} {Mathematical foundations for a compositional distributed
  model of meaning}.{\BBCQ}
\newblock
\APACjournalVolNumPages{Linguistic Analysis}{36}{}{345--384}.
\PrintBackRefs{\CurrentBib}

\bibitem [\protect \citeauthoryear {%
Deutsch%
}{%
Deutsch%
}{%
{\protect \APACyear {1999}}%
}]{%
Deutsch_1999}
\APACinsertmetastar {%
Deutsch_1999}%
\begin{APACrefauthors}%
Deutsch, D.%
\end{APACrefauthors}%
\unskip\
\newblock
\APACrefYearMonthDay{1999}{}{}.
\newblock
{\BBOQ}\APACrefatitle {Quantum Theory of Probability and Decisions} {Quantum
  theory of probability and decisions}.{\BBCQ}
\newblock
\BIn{} \APACrefbtitle {Proceedings of the {R}oyal {S}ociety of {L}ondon {A455}}
  {Proceedings of the {R}oyal {S}ociety of {L}ondon {A455}}\ (\BPGS\
  3129--3137).
\PrintBackRefs{\CurrentBib}

\bibitem [\protect \citeauthoryear {%
Eisenberg%
\ \BBA {} Barry%
}{%
Eisenberg%
\ \BBA {} Barry%
}{%
{\protect \APACyear {1988}}%
}]{%
Eisenberg_etal1988}
\APACinsertmetastar {%
Eisenberg_etal1988}%
\begin{APACrefauthors}%
Eisenberg, M.%
\BCBT {}\ \BBA {} Barry, C.%
\end{APACrefauthors}%
\unskip\
\newblock
\APACrefYearMonthDay{1988}{}{}.
\newblock
{\BBOQ}\APACrefatitle {Order Effects: A Study of the Possible Influence of
  Presentation Order on User Judgments of Document Relevance} {Order effects: A
  study of the possible influence of presentation order on user judgments of
  document relevance}.{\BBCQ}
\newblock
\APACjournalVolNumPages{Journal of the American Society for Information
  Science}{39}{5}{293--300}.
\PrintBackRefs{\CurrentBib}

\bibitem [\protect \citeauthoryear {%
Frommholz%
\ \protect \BOthers {.}}{%
Frommholz%
\ \protect \BOthers {.}}{%
{\protect \APACyear {2010}}%
}]{%
Ingo-Polyrepresentation}
\APACinsertmetastar {%
Ingo-Polyrepresentation}%
\begin{APACrefauthors}%
Frommholz, I.%
, Larsen, B.%
, Piwowarski, B.%
, Lalmas, M.%
, Ingwersen, P.%
\BCBL {}\ \BBA {} van Rijsbergen, C\BPBI J.%
\end{APACrefauthors}%
\unskip\
\newblock
\APACrefYearMonthDay{2010}{}{}.
\newblock
{\BBOQ}\APACrefatitle {Supporting Polyrepresentation in a Quantum-inspired
  Geometrical Retrieval Framework} {Supporting polyrepresentation in a
  quantum-inspired geometrical retrieval framework}.{\BBCQ}
\newblock
\BIn{} \APACrefbtitle {{Proceedings of the 3rd Information Interaction in
  Context symposium}} {{Proceedings of the 3rd Information Interaction in
  Context symposium}}\ (\BPGS\ 115--124).
\PrintBackRefs{\CurrentBib}

\bibitem [\protect \citeauthoryear {%
Harrigan%
\ \BBA {} Rudolph%
}{%
Harrigan%
\ \BBA {} Rudolph%
}{%
{\protect \APACyear {2007}}%
}]{%
Harrigan_Terry_2007}
\APACinsertmetastar {%
Harrigan_Terry_2007}%
\begin{APACrefauthors}%
Harrigan, N.%
\BCBT {}\ \BBA {} Rudolph, T.%
\end{APACrefauthors}%
\unskip\
\newblock
\APACrefYearMonthDay{2007}{}{}.
\newblock
\APACrefbtitle {{Ontological models and the interpretation of contextuality}.}
  {{Ontological models and the interpretation of contextuality}.}
\newblock
\APAChowpublished {arXiv:0709.4266v1}.
\PrintBackRefs{\CurrentBib}

\bibitem [\protect \citeauthoryear {%
Harrigan%
\ \BBA {} Spekkens%
}{%
Harrigan%
\ \BBA {} Spekkens%
}{%
{\protect \APACyear {2010}}%
}]{%
Harrigan_Spekkens_2010}
\APACinsertmetastar {%
Harrigan_Spekkens_2010}%
\begin{APACrefauthors}%
Harrigan, N.%
\BCBT {}\ \BBA {} Spekkens, R\BPBI W.%
\end{APACrefauthors}%
\unskip\
\newblock
\APACrefYearMonthDay{2010}{}{}.
\newblock
{\BBOQ}\APACrefatitle {{Einstein, incompleteness, and the epistemic view of
  quantum states}} {{Einstein, incompleteness, and the epistemic view of
  quantum states}}.{\BBCQ}
\newblock
\APACjournalVolNumPages{Foundations of Physics}{40}{2}{125--157}.
\PrintBackRefs{\CurrentBib}

\bibitem [\protect \citeauthoryear {%
Huertas-Rosero%
, Azzopardi%
\BCBL {}\ \BBA {} van Rijsbergen%
}{%
Huertas-Rosero%
\ \protect \BOthers {.}}{%
{\protect \APACyear {2008}}%
}]{%
Alvaro_etal2008}
\APACinsertmetastar {%
Alvaro_etal2008}%
\begin{APACrefauthors}%
Huertas-Rosero, A\BPBI F.%
, Azzopardi, L.%
\BCBL {}\ \BBA {} van Rijsbergen, C\BPBI J.%
\end{APACrefauthors}%
\unskip\
\newblock
\APACrefYearMonthDay{2008}{}{}.
\newblock
{\BBOQ}\APACrefatitle {Characterising through Erasing: A Theoretical Framework
  for Representing Documents Inspired by Quantum Theory} {Characterising
  through erasing: A theoretical framework for representing documents inspired
  by quantum theory}.{\BBCQ}
\newblock
\BIn{} \APACrefbtitle {{Proceedings of the 2nd AAAI Quantum Interaction
  Symposium}} {{Proceedings of the 2nd AAAI Quantum Interaction Symposium}}\
  (\BPGS\ 160--163).
\PrintBackRefs{\CurrentBib}

\bibitem [\protect \citeauthoryear {%
Kantor%
}{%
Kantor%
}{%
{\protect \APACyear {2007}}%
}]{%
Kantor_2007}
\APACinsertmetastar {%
Kantor_2007}%
\begin{APACrefauthors}%
Kantor, P\BPBI B.%
\end{APACrefauthors}%
\unskip\
\newblock
\APACrefYearMonthDay{2007}{}{}.
\newblock
{\BBOQ}\APACrefatitle {{Keith van Rijsbergen}, The Geometry of Information
  Retrieval} {{Keith van Rijsbergen}, the geometry of information
  retrieval}.{\BBCQ}
\newblock
\APACjournalVolNumPages{Information Retrieval}{10}{4-5}{485-489}.
\PrintBackRefs{\CurrentBib}

\bibitem [\protect \citeauthoryear {%
Kirkpatrick%
}{%
Kirkpatrick%
}{%
{\protect \APACyear {2003}}%
}]{%
kirkpatrick-2003}
\APACinsertmetastar {%
kirkpatrick-2003}%
\begin{APACrefauthors}%
Kirkpatrick, K\BPBI A.%
\end{APACrefauthors}%
\unskip\
\newblock
\APACrefYearMonthDay{2003}{}{}.
\newblock
{\BBOQ}\APACrefatitle {{"Quantal"} behavior in classical probability}
  {{"Quantal"} behavior in classical probability}.{\BBCQ}
\newblock
\APACjournalVolNumPages{Foundations of Physics Letters}{16}{3}{199-224}.
\PrintBackRefs{\CurrentBib}

\bibitem [\protect \citeauthoryear {%
Law%
, Klobu\v{c}ar%
\BCBL {}\ \BBA {} Pipan%
}{%
Law%
\ \protect \BOthers {.}}{%
{\protect \APACyear {2006}}%
}]{%
Law_etal_2006}
\APACinsertmetastar {%
Law_etal_2006}%
\begin{APACrefauthors}%
Law, E\BPBI L.%
, Klobu\v{c}ar, T.%
\BCBL {}\ \BBA {} Pipan, M.%
\end{APACrefauthors}%
\unskip\
\newblock
\APACrefYearMonthDay{2006}{}{}.
\newblock
{\BBOQ}\APACrefatitle {{User Effect in Evaluating Personalized Information
  Retrieval Systems}} {{User Effect in Evaluating Personalized Information
  Retrieval Systems}}.{\BBCQ}
\newblock
\BIn{} \APACrefbtitle {{Proceedings of the 1st European conference on
  Technology Enhanced Learning: Innovative Approaches for Learning and
  Knowledge Sharing}} {{Proceedings of the 1st European conference on
  Technology Enhanced Learning: Innovative Approaches for Learning and
  Knowledge Sharing}}\ (\BPGS\ 257--271).
\PrintBackRefs{\CurrentBib}

\bibitem [\protect \citeauthoryear {%
Leong%
, Vetere%
\BCBL {}\ \BBA {} Howard%
}{%
Leong%
\ \protect \BOthers {.}}{%
{\protect \APACyear {2006}}%
}]{%
Leong_etal2006}
\APACinsertmetastar {%
Leong_etal2006}%
\begin{APACrefauthors}%
Leong, T\BPBI W.%
, Vetere, F.%
\BCBL {}\ \BBA {} Howard, S.%
\end{APACrefauthors}%
\unskip\
\newblock
\APACrefYearMonthDay{2006}{}{}.
\newblock
{\BBOQ}\APACrefatitle {Randomness as a resource for design} {Randomness as a
  resource for design}.{\BBCQ}
\newblock
\BIn{} \APACrefbtitle {{Proceedings of the 6th ACM conference on Designing
  Interactive systems}} {{Proceedings of the 6th ACM conference on Designing
  Interactive systems}}\ (\BPGS\ 132--139).
\PrintBackRefs{\CurrentBib}

\bibitem [\protect \citeauthoryear {%
Melucci%
}{%
Melucci%
}{%
{\protect \APACyear {2008}}%
}]{%
Melucci_08}
\APACinsertmetastar {%
Melucci_08}%
\begin{APACrefauthors}%
Melucci, M.%
\end{APACrefauthors}%
\unskip\
\newblock
\APACrefYearMonthDay{2008}{}{}.
\newblock
{\BBOQ}\APACrefatitle {A basis for information retrieval in context} {A basis
  for information retrieval in context}.{\BBCQ}
\newblock
\APACjournalVolNumPages{ACM Transactions on Information Systems}{26}{3}{1-41}.
\PrintBackRefs{\CurrentBib}

\bibitem [\protect \citeauthoryear {%
Melucci%
\ \BBA {} van Rijsbergen%
}{%
Melucci%
\ \BBA {} van Rijsbergen%
}{%
{\protect \APACyear {2011}}%
}]{%
Melucci_Keith_2011}
\APACinsertmetastar {%
Melucci_Keith_2011}%
\begin{APACrefauthors}%
Melucci, M.%
\BCBT {}\ \BBA {} van Rijsbergen, C\BPBI J.%
\end{APACrefauthors}%
\unskip\
\newblock
\APACrefYearMonthDay{2011}{}{}.
\newblock
{\BBOQ}\APACrefatitle {Quantum Mechanics and Information Retrieval} {Quantum
  mechanics and information retrieval}.{\BBCQ}
\newblock
\BIn{} M.~Melucci, R.~Baeza-Yates\BCBL {}\ \BBA {} W\BPBI B.~Croft\ (\BEDS),
  \APACrefbtitle {Advanced Topics in Information Retrieval} {Advanced topics in
  information retrieval}\ (\BVOL~33, \BPGS\ 125--155).
\newblock
\APACaddressPublisher{}{Heidelberg:Springer}.
\PrintBackRefs{\CurrentBib}

\bibitem [\protect \citeauthoryear {%
Pavlovic%
}{%
Pavlovic%
}{%
{\protect \APACyear {2008}}%
}]{%
PavlovicD_QI08}
\APACinsertmetastar {%
PavlovicD_QI08}%
\begin{APACrefauthors}%
Pavlovic, D.%
\end{APACrefauthors}%
\unskip\
\newblock
\APACrefYearMonthDay{2008}{}{}.
\newblock
{\BBOQ}\APACrefatitle {On quantum statistics in data analysis} {On quantum
  statistics in data analysis}.{\BBCQ}
\newblock
\BIn{} \APACrefbtitle {{Proceedings of the 2nd AAAI Quantum Interaction
  Symposium}} {{Proceedings of the 2nd AAAI Quantum Interaction Symposium}}\
  (\BPGS\ 260--267).
\PrintBackRefs{\CurrentBib}

\bibitem [\protect \citeauthoryear {%
Piwowarski%
, Frommholz%
, Lalmas%
\BCBL {}\ \BBA {} van Rijsbergen%
}{%
Piwowarski%
\ \protect \BOthers {.}}{%
{\protect \APACyear {2010}}%
}]{%
Piwowarski_what_2010}
\APACinsertmetastar {%
Piwowarski_what_2010}%
\begin{APACrefauthors}%
Piwowarski, B.%
, Frommholz, I.%
, Lalmas, M.%
\BCBL {}\ \BBA {} van Rijsbergen, C\BPBI J.%
\end{APACrefauthors}%
\unskip\
\newblock
\APACrefYearMonthDay{2010}{}{}.
\newblock
{\BBOQ}\APACrefatitle {What can quantum theory bring to information retrieval}
  {What can quantum theory bring to information retrieval}.{\BBCQ}
\newblock
\BIn{} \APACrefbtitle {{Proceedings of the 19th {ACM} International Conference
  on Information and Knowledge Management}} {{Proceedings of the 19th {ACM}
  International Conference on Information and Knowledge Management}}\ (\BPGS\
  59--68).
\PrintBackRefs{\CurrentBib}

\bibitem [\protect \citeauthoryear {%
Piwowarski%
\ \BBA {} Lalmas%
}{%
Piwowarski%
\ \BBA {} Lalmas%
}{%
{\protect \APACyear {2009}}%
}]{%
Piwowarski_etal_2009}
\APACinsertmetastar {%
Piwowarski_etal_2009}%
\begin{APACrefauthors}%
Piwowarski, B.%
\BCBT {}\ \BBA {} Lalmas, M.%
\end{APACrefauthors}%
\unskip\
\newblock
\APACrefYearMonthDay{2009}{}{}.
\newblock
{\BBOQ}\APACrefatitle {A Quantum-based Model for Interactive Information
  Retrieval} {A quantum-based model for interactive information
  retrieval}.{\BBCQ}
\newblock
\BIn{} \APACrefbtitle {{Proceedings of the 2nd International Conference on the
  Theory of Information Retrieval}} {{Proceedings of the 2nd International
  Conference on the Theory of Information Retrieval}}\ (\BPGS\ 224--231).
\newblock
\APACaddressPublisher{}{Springer}.
\PrintBackRefs{\CurrentBib}

\bibitem [\protect \citeauthoryear {%
Randall%
\ \BBA {} Foulis%
}{%
Randall%
\ \BBA {} Foulis%
}{%
{\protect \APACyear {1979}}%
}]{%
Randall_1979}
\APACinsertmetastar {%
Randall_1979}%
\begin{APACrefauthors}%
Randall, C.%
\BCBT {}\ \BBA {} Foulis, D.%
\end{APACrefauthors}%
\unskip\
\newblock
\APACrefYearMonthDay{1979}{}{}.
\newblock
{\BBOQ}\APACrefatitle {Tensor products of quantum logics do not exist} {Tensor
  products of quantum logics do not exist}.{\BBCQ}
\newblock
\APACjournalVolNumPages{Notices of the American Mathematical
  Society}{26}{6}{557}.
\PrintBackRefs{\CurrentBib}

\bibitem [\protect \citeauthoryear {%
Rieffel%
}{%
Rieffel%
}{%
{\protect \APACyear {2007}}%
}]{%
Rieffel_2007}
\APACinsertmetastar {%
Rieffel_2007}%
\begin{APACrefauthors}%
Rieffel, E.%
\end{APACrefauthors}%
\unskip\
\newblock
\APACrefYearMonthDay{2007}{}{}.
\newblock
{\BBOQ}\APACrefatitle {Certainty and Uncertainty in Quantum Information
  Processing} {Certainty and uncertainty in quantum information
  processing}.{\BBCQ}
\newblock
\BIn{} \APACrefbtitle {{Proceedings of the AAAI Spring Symposium on Quantum
  Interaction}} {{Proceedings of the AAAI Spring Symposium on Quantum
  Interaction}}\ (\BPGS\ 134--141).
\PrintBackRefs{\CurrentBib}

\bibitem [\protect \citeauthoryear {%
Spekkens%
}{%
Spekkens%
}{%
{\protect \APACyear {2007}}%
}]{%
toy_2007}
\APACinsertmetastar {%
toy_2007}%
\begin{APACrefauthors}%
Spekkens, R\BPBI W.%
\end{APACrefauthors}%
\unskip\
\newblock
\APACrefYearMonthDay{2007}{}{}.
\newblock
{\BBOQ}\APACrefatitle {In defense of the epistemic view of quantum states: a
  toy theory} {In defense of the epistemic view of quantum states: a toy
  theory}.{\BBCQ}
\newblock
\APACjournalVolNumPages{Physical Review A}{75}{}{032110}.
\PrintBackRefs{\CurrentBib}

\bibitem [\protect \citeauthoryear {%
Valentini%
}{%
Valentini%
}{%
{\protect \APACyear {2009}}%
}]{%
Valentini_2009}
\APACinsertmetastar {%
Valentini_2009}%
\begin{APACrefauthors}%
Valentini, A.%
\end{APACrefauthors}%
\unskip\
\newblock
\APACrefYearMonthDay{2009}{}{}.
\newblock
\APACrefbtitle {Beyond the Quantum.} {Beyond the quantum.}
\newblock
\APAChowpublished {Physics World, 32-37}.
\PrintBackRefs{\CurrentBib}

\bibitem [\protect \citeauthoryear {%
van Rijsbergen%
}{%
van Rijsbergen%
}{%
{\protect \APACyear {1981}}%
}]{%
Rijsbergen_experiment}
\APACinsertmetastar {%
Rijsbergen_experiment}%
\begin{APACrefauthors}%
van Rijsbergen, C\BPBI J.%
\end{APACrefauthors}%
\unskip\
\newblock
\APACrefYearMonthDay{1981}{}{}.
\newblock
\APACrefbtitle {Retrieval effectiveness.} {Retrieval effectiveness.}
\newblock
\APAChowpublished {{In K. Sp\"arck Jones~(Eds.), Information retrieval
  experiment (pp.32--43). London:Butterworths}}.
\PrintBackRefs{\CurrentBib}

\bibitem [\protect \citeauthoryear {%
van Rijsbergen%
}{%
van Rijsbergen%
}{%
{\protect \APACyear {1986}}%
}]{%
Keith_nonclassic_1986}
\APACinsertmetastar {%
Keith_nonclassic_1986}%
\begin{APACrefauthors}%
van Rijsbergen, C\BPBI J.%
\end{APACrefauthors}%
\unskip\
\newblock
\APACrefYearMonthDay{1986}{}{}.
\newblock
{\BBOQ}\APACrefatitle {A Non-Classical Logic for Information Retrieval} {A
  non-classical logic for information retrieval}.{\BBCQ}
\newblock
\APACjournalVolNumPages{Computer Journal}{29}{6}{481-485}.
\PrintBackRefs{\CurrentBib}

\bibitem [\protect \citeauthoryear {%
van Rijsbergen%
}{%
van Rijsbergen%
}{%
{\protect \APACyear {1989}}%
}]{%
Keith_towardslogic}
\APACinsertmetastar {%
Keith_towardslogic}%
\begin{APACrefauthors}%
van Rijsbergen, C\BPBI J.%
\end{APACrefauthors}%
\unskip\
\newblock
\APACrefYearMonthDay{1989}{}{}.
\newblock
{\BBOQ}\APACrefatitle {Towards an information logic} {Towards an information
  logic}.{\BBCQ}
\newblock
\APACjournalVolNumPages{SIGIR Forum}{23}{}{77--86}.
\PrintBackRefs{\CurrentBib}

\bibitem [\protect \citeauthoryear {%
van Rijsbergen%
}{%
van Rijsbergen%
}{%
{\protect \APACyear {1996}}%
}]{%
Keith_qlogic_1996}
\APACinsertmetastar {%
Keith_qlogic_1996}%
\begin{APACrefauthors}%
van Rijsbergen, C\BPBI J.%
\end{APACrefauthors}%
\unskip\
\newblock
\APACrefYearMonthDay{1996}{}{}.
\newblock
{\BBOQ}\APACrefatitle {Quantum Logic and Information Retrieval} {Quantum logic
  and information retrieval}.{\BBCQ}
\newblock
\BIn{} \APACrefbtitle {{Proceedings of Workshop on Logical and Uncertainty
  Models in Information Retrieval, University of Glasgow, Glasgow, July, 1-2.}}
  {{Proceedings of Workshop on Logical and Uncertainty Models in Information
  Retrieval, University of Glasgow, Glasgow, July, 1-2.}}
\PrintBackRefs{\CurrentBib}

\bibitem [\protect \citeauthoryear {%
van Rijsbergen%
}{%
van Rijsbergen%
}{%
{\protect \APACyear {2004}}%
}]{%
Rijsbergen_gir}
\APACinsertmetastar {%
Rijsbergen_gir}%
\begin{APACrefauthors}%
van Rijsbergen, C\BPBI J.%
\end{APACrefauthors}%
\unskip\
\newblock
\APACrefYear{2004}.
\newblock
\APACrefbtitle {The Geometry of Information Retrieval} {The geometry of
  information retrieval}.
\newblock
\APACaddressPublisher{}{Cambridge University Press}.
\PrintBackRefs{\CurrentBib}

\bibitem [\protect \citeauthoryear {%
van Rijsbergen%
}{%
van Rijsbergen%
}{%
{\protect \APACyear {2006}}%
}]{%
Rijsbergen_keynote_sigir}
\APACinsertmetastar {%
Rijsbergen_keynote_sigir}%
\begin{APACrefauthors}%
van Rijsbergen, C\BPBI J.%
\end{APACrefauthors}%
\unskip\
\newblock
\APACrefYearMonthDay{2006}{}{}.
\newblock
{\BBOQ}\APACrefatitle {{Quantum haystacks}} {{Quantum haystacks}}.{\BBCQ}
\newblock
\BIn{} \APACrefbtitle {{Proceedings of the 29th annual International ACM SIGIR
  Conference on Research and Development in Information Retrieval}}
  {{Proceedings of the 29th annual International ACM SIGIR Conference on
  Research and Development in Information Retrieval}}\ (\BPGS\ 1--2).
\PrintBackRefs{\CurrentBib}

\bibitem [\protect \citeauthoryear {%
Wang%
, Song%
, Zhang%
, Hou%
\BCBL {}\ \BBA {} Bruza%
}{%
Wang%
\ \protect \BOthers {.}}{%
{\protect \APACyear {2010}}%
}]{%
Wang_etal_2010}
\APACinsertmetastar {%
Wang_etal_2010}%
\begin{APACrefauthors}%
Wang, J.%
, Song, D.%
, Zhang, P.%
, Hou, Y.%
\BCBL {}\ \BBA {} Bruza, P.%
\end{APACrefauthors}%
\unskip\
\newblock
\APACrefYearMonthDay{2010}{}{}.
\newblock
{\BBOQ}\APACrefatitle {Explanation of Relevance Judgement Discrepancy with
  Quantum Interference} {Explanation of relevance judgement discrepancy with
  quantum interference}.{\BBCQ}
\newblock
\BIn{} \APACrefbtitle {{AAAI-Fall 2010 Symposium on Quantum Informatics for
  Cognitive, Social, and Semantic Processes (QI)}} {{AAAI-Fall 2010 Symposium
  on Quantum Informatics for Cognitive, Social, and Semantic Processes (QI)}}\
  (\BPGS\ 117--124).
\PrintBackRefs{\CurrentBib}

\bibitem [\protect \citeauthoryear {%
Zuccon%
\ \BBA {} Azzopardi%
}{%
Zuccon%
\ \BBA {} Azzopardi%
}{%
{\protect \APACyear {2010}}%
}]{%
Guido_ecir2010}
\APACinsertmetastar {%
Guido_ecir2010}%
\begin{APACrefauthors}%
Zuccon, G.%
\BCBT {}\ \BBA {} Azzopardi, L.%
\end{APACrefauthors}%
\unskip\
\newblock
\APACrefYearMonthDay{2010}{}{}.
\newblock
{\BBOQ}\APACrefatitle {Using the Quantum Probability Ranking Principle to Rank
  Interdependent Documents} {Using the quantum probability ranking principle to
  rank interdependent documents}.{\BBCQ}
\newblock
\BIn{} \APACrefbtitle {{Proceedings of the 32nd European Conference on
  Information Retrieval~(ECIR)}} {{Proceedings of the 32nd European Conference
  on Information Retrieval~(ECIR)}}\ (\BPGS\ 357--369).
\PrintBackRefs{\CurrentBib}

\end{thebibliography}

\appendix

\section{Reading the specific entries of the state vector would imply ``magical'' powers}
\label{appendix-qi-proof}
The aim of this illustration is to show why the ability to read the specific entries of the state vector enables one to perform things that would be considered ``magical" within the framework of generic quantum physics.
If documents or queries are represented as a vector of complex numbers on a classical computer then we can read out any element of the state vector. This is different from a genuinely quantum representation of a quantum state, in which 
we destroy the quantum state in so doing.
We provide merely an illustrative example of a problem believed to be extremely hard~(NP-complete).

The Hamiltonian cycle problem is a decision problem of resolving the question of whether a given graph has a Hamiltonian cycle, where a Hamiltonian cycle is a path which visits each vertex 1..n exactly once, returning to the original vertex in the final step. Classically it is difficult to find such a Hamiltonian cycle in a given graph, meaning that no efficient algorithm~(polynomial time) exists to solve this problem. However, one can efficiently verify whether a potential cycle is Hamiltonian. Just try it on the graph! Taking benefit of this ability and of course few quantum tricks, a quantum computational model is able to perform such verification for all possible paths on a given graph simultaneously.
However, only if we can read the entries of the state vector then we can eventually use this to solve this problem efficiently. QM does not let us to do this which is why quantum computers can not solve this problem efficiently.

\vspace{2mm}
\noindent
\textbf{How does one prepare and store all possible paths as input to this model?}
\vspace{2mm}

As we described before the state of $n$ two-level quantum systems~(quantum bits or qubits) is described by $2^n$ complex numbers where each complex number reflects one possible result of measuring the collection of $n$ qubits. If you prefer you can imagine that $2^n$ complex numbers are stored somehow in $n$ qubits, where in our example $n$ corresponds to the number of vertices in the graph. However, the meaning of storing information is different than what it generally means in classical computing. The generic state of these $n$ qubits is known to be in a superposition of these $2^n$ possible states. See Section~\ref{sec:superposition} for further details on quantum superposition. Such a superposed state can be prepared through performing a small number of operations in the lab.

\vspace{2mm}
\noindent
\textbf{What to do with the superposition of all possible paths?}
\vspace{2mm}

Let us assume that we can design a function $f$ with respect to the structure of the graph, which takes as input one possible path on graph $G$ and verifies if the input is a Hamiltonian cycle, returning 1 or 0. Let $x$ corresponds to a possible path on the graph, i.e. one of many possible permutation of vertices. If $\exists x, f(x)=1$, then the given graph has a Hamiltonian cycle. Obviously if $\forall x, f(x)=0$ then no Hamiltonian cycle exists in $G$.
Given that we can construct a sequence of reversible transformations~($U_{f}$) to compute function $f$, the quantum circuit implementing the
above takes as input a superposition of all the possible inputs~(n input qubits), and stores the output in an answer qubit.

\vspace{2mm}
\noindent
\textbf{Can one read the result for an arbitrary path?}
\vspace{2mm}

The above procedure may leave one to imagine that this quantum circuit calculates the outcome of function $f$ for all possible paths simultaneously. This looks beneficial as we are able to verify for each possible cycle if it is a Hamiltonian cycle or not. However the difficulty arises as the final state of the system is a superposed state which means we need to do measurement to be aware of the outcome for each path. By measuring the final state, we are able to know the value of $f(x)$ for just one random value of $x$ as the state will collapse and we will loose all the information. Is there anyway to unleash this inherent information?

\vspace{2mm}
\noindent 
\textbf{Is there any possibility to extract the desired output?} 
\vspace{2mm}

In this example we aim to know whether a given graph has a Hamiltonian cycle, i.e. if $\exists x, f(x)=1$. 
For this purpose, we are able to recombine the states in the final superposition state in a way to push the first $n$ qubits to the zero state and leave the final answer in the answer qubit. This recombination is possible, using a small number of gates in the lab, taking the benefits of quantum interference~(See Section~\ref{interference} for more details). It looks as if we are re-arranging the amplitudes between those states in a way to boost the probability of getting the desired output. The final answer qubit will contain the result in the following way. If no such Hamiltonian cycle exists in the given graph, it ends up in state $|0\rangle$, where its state vector corresponds to $\begin{pmatrix} 1 \\ 0 \end{pmatrix}$; if there is a Hamiltonian cycle, the final state is in a superposition of states $|0\rangle$ and $|1\rangle$ with amplitudes $\sqrt{1-\frac{1}{2^n}}$ and $\sqrt{\frac{1}{2^n}}$ respectively and the corresponding state vector is
$\begin{pmatrix} \sqrt{1-\frac{1}{2^n}} \\ \sqrt{\frac{1}{2^n}}\end{pmatrix}$. 

The above recombination reduce the dimension of the space to two, which means we merely required to examine one qubit to know the answer. The bad news is that in the real physical world we cannot look at this state vector. We need to perform a measurement and due to the small probability of the $|1\rangle$ state, the chance of observing this desired configuration is very low~($\frac{1}{2^n}$), and we have to repeat the experiments many times.
Imagine that we could open the state vector to look and read out any desired entry of it: we would be able to answer the graph Hamiltonian cycle problem in one evaluation of $U_{f}$, which seems magical as opposed to doing these comparisons in a classical computer. However, as it is well-known the rules of QM forbid this observation.

\end{document}